\documentclass[conference]{IEEEtran}
\IEEEoverridecommandlockouts
\usepackage{amsmath,amssymb,amsfonts}
\usepackage{algorithmic}
\usepackage{graphicx}
\usepackage{textcomp}
\usepackage{xcolor}
\usepackage{todonotes}
\usepackage{soul}
\usepackage{verbatim}
\usepackage{fancyvrb}
\usepackage{hyperref}
\usepackage{multirow}
\usepackage{float}
\hypersetup{colorlinks=true,linkcolor=black,urlcolor=blue, citecolor=black}

\def\BibTeX{{\rm B\kern-.05em{\sc i\kern-.025em b}\kern-.08em
    T\kern-.1667em\lower.7ex\hbox{E}\kern-.125emX}}

\newcommand{\ps}{PowerShell}
\DeclareUnicodeCharacter{2212}{-}

\begin{document}
\title{Command-line Obfuscation Detection using Small Language Models\\
}

\author{\IEEEauthorblockN{1\textsuperscript{st} Vojtech Outrata}
\IEEEauthorblockA{\textit{Cisco systems} \\
Prague, Czechia \\
voutrata@cisco.com}
\and
\IEEEauthorblockN{2\textsuperscript{nd} Michael Adam Polak}
\IEEEauthorblockA{\textit{Cisco systems} \\
Prague, Czechia \\
mipolak@cisco.com}
\and
\IEEEauthorblockN{3\textsuperscript{rd} Martin Kopp}
\IEEEauthorblockA{\textit{Cisco systems} \\
Prague, Czechia \\
markopp@cisco.com}
}

\maketitle

\begin{abstract}

To avoid detection, adversaries often use command-line obfuscation. There are numerous techniques of the command-line obfuscation, all designed to alter the command-line syntax without affecting its original functionality. This variability forces most security solutions to create an exhaustive enumeration of signatures for even a single pattern.
In contrast to using signatures, we have implemented a scalable NLP-based detection method that leverages a custom-trained, small transformer language model that can be applied to any source of execution logs. The evaluation on top of real-world telemetry demonstrates that our approach yields high-precision detections even on high-volume telemetry from a diverse set of environments spanning from universities and businesses to healthcare or finance. 
The practical value is demonstrated in a case study of real-world samples detected by our model. We show the model's superiority to signatures on established malware known to employ obfuscation and showcase previously unseen obfuscated samples detected by our model. 


\end{abstract}

\section{Introduction}

Traditional malware detection approaches include matching hashes of executed binaries or writing signatures for the malicious command-lines. To avoid detection by hash matching, attackers leverage common binaries already preinstalled on the targeted machine, so-called living off the land binaries (LoLBins). These binaries are used for everyday activities, so matching their hash is not an option.

When LOLBins are utilized, signature matching on top of the command-line is one of the few available options for detecting stealthy malware. Similarly, as with hash matching, attackers try to avoid signatures as well. They change the case of the characters, add unnecessary white-space characters, or add characters ignored by the command-line interpreter such as `` \^{} " or `` \`{} ", etc. Collectively, these modifications that only change the syntax of the command and not the execution are called obfuscation techniques. Writing signatures for obfuscation techniques is a prohibitively difficult task since there are endless possibilities when combining the techniques. This complexity makes it impossible for signatures to capture the generic obfuscation patterns and requires a more sophisticated detection approach.

In our work, we focus on obfuscation detection on top of general command-line execution logs, specifically on top of LOLBin executions. Since the command-line resembles a language where each command represents a sentence, we propose a transformer-based language model for obfuscation detection. As the model is developed from scratch, the paper describes each step of the training process and a thorough evaluation of the model on real-world telemetry with an emphasis on malicious activity.

To demonstrate the capabilities of our detection method, we present a case study with commands for existing malware that avoided traditional detection techniques and previously unseen samples of obfuscated malicious commands all detected by our model.

\begin{figure*}
    \centering
    \includegraphics[width=0.99\textwidth]{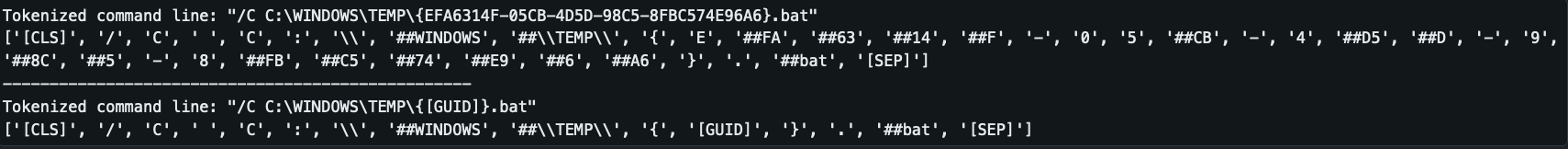}
    \caption{Tokenization of raw command-line compared to tokenization of preprocessed command-line by the same tokenizer.  The GUID pattern in the raw command-line is tokenized into many non-meaningful tokens. In contrast, a single general token represents the pattern in the tokenized preprocessed command-line. }
    \label{fig:preprocessing_tokenization}
\end{figure*}

\section{Prior Art}

To address the limitations of signature-based detectors, extensive efforts have been directed toward developing machine-learning solutions for specific obfuscation and malware detection scenarios. One such effort resulted in the creation of a specialized detector for obfuscation in Java malware~\cite{java_obfuscation}. As the model is tailored specifically for Java malware, it utilizes Java bytecode itself for feature extraction. A similar approach is described in \cite{android_obfuscation}, where authors apply a machine-learning approach to detect obfuscation techniques in Android applications. The authors use the application opcodes (instruction machine code) to detect obfuscated applications in two ways. The first approach performs feature extraction followed by a binary classifier, while the second approach utilizes LSTM-RNN neural networks on the opcodes directly. A more general approach is studied in \cite{malware_classification}, where the authors apply similar methods on top of system logs produced by running the Microsoft anti-malware engine on top of the Windows portable executable (PE) file format. 

All previously described methods differ significantly from our approach as they specialize in in-depth analysis of the bytecode of applications or even directly running the executable and recording its activity. Our work targets obfuscation detection at a higher level of the command-line interface and text analysis. Such an approach is studied in~\cite{malicious_PowerShell}, where the authors target the classification of malicious \ps{} commands using various NLP techniques, including convolutional/recurrent neural networks. However, obfuscation techniques are only mentioned as one possible feature of malicious samples, and the work focuses on \ps{} only. A detection approach targeting specifically obfuscation for \ps{} has been developed by Daniel Bohanon's project, Revoke-Obfuscation \cite{revoke_obfuscation}. Their method relies on extracting features from \ps{'s} inbuilt Abstract Syntax Tree followed by a shallow classifier. While a powerful model, it can only detect \ps{} obfuscation and is not scalable to large-scale deployment due to the expensive construction of the syntax tree followed by feature extraction.

We build upon previous works in two main aspects. Firstly by leveraging the transformer-based architecture proven superior to other natural language processing methods in recent years. As obfuscation techniques greatly impact the structure of the command-line, the transformer model easily recognizes obfuscated samples. Consequently, smaller models scalable to high-volume telemetries can be utilized without loss of classification performance. Secondly, we were able to extend the model's detection capabilities beyond \ps{} to numerous other LOLBins.

\section{Method}
\label{sec:method}

Developing a transformer-based language model from scratch requires multiple data preprocessing and model training steps. The process can be separated into the following phases:
\begin{enumerate}
    \item Data preprocessing
    \item Tokenizer training
    \item Model pre-training
    \item Model fine-tuning
\end{enumerate}
We provide a detailed description of each step in its respective section.

\subsection{Data preprocessing}
Transformer-based language models comprehend text by dividing it into smaller units (tokens) and assessing the meaning of each token in relation to its context, such as the tokens in the rest of the sentence. For this mechanism to yield effective results, the tokens should convey relevant meaning. In command-line data, however, many artifacts exist that are not relevant to other parts of the command or the task at hand.

We have identified that concrete values of specific patterns, such as IP addresses, dates, or GUIDs in the command-lines, are irrelevant for obfuscation detection. We replace these non-meaningful command-line parts with a generic token in the preprocessing phase to avoid feeding them into the model. All such patterns and their substitute tokens are listed in Table~\ref{data_preprocessing}.

For example, the GUID in the following cmd command-line:
\begin{verbatim}
    /C C:\WINDOWS\TEMP\{EFA6314F-
    05CB-4D5D-98C5-8FBC574E96A6}.bat
\end{verbatim}
is replaced with a generic token [GUID], resulting in:
\begin{verbatim}
    /C C:\WINDOWS\TEMP\{[GUID]}.bat
\end{verbatim}
The preprocessed command-lines are more straightforward to tokenize, as demonstrated in Figure~\ref{fig:preprocessing_tokenization}. 
In the raw command, the GUID pattern is split into numerous short tokens, resulting in a significantly longer token sequence. In comparison, the preprocessed command-line contains only a single [GUID] token representing the entire pattern.

In cases where these patterns are obfuscated (e.g., Raspberry Robin malware), pattern matching will fail; consequently, the obfuscated parts will not be replaced.

\begin{table}[!t]
\renewcommand{\arraystretch}{1.3}
\caption{Patterns replaced in the command-line}
\label{data_preprocessing}
\centering
\begin{tabular}{|c||c|}
\hline
\bfseries Replaced pattern & \bfseries Token\\
\hline\hline
GUID (globally unique identifier) & [GUID]\\ \hline
IP (both v4 and v6) & [IP]\\ \hline
Date (numerous formats) & [DATE]\\ \hline
Number & [NUM]\\ \hline
URL & [URL]\\
\hline
\end{tabular}
\end{table}

\subsection{Tokenizer training}
Before the preprocessed data can be fed into the model, the text has to be split into smaller tokens by a trained tokenizer. One feasible approach is to adapt an already existing tokenizer, such as the tokenizer from BERT~\cite{bert}. However, most pre-trained tokenizers (along with the model) are trained on natural language. The command-lines significantly differ from natural language in both syntax and semantics. Since the tokenizer quality heavily influences downstream task performance and the speed of the resulting model, we train a custom tokenizer for command-line logs (presented in Section~\ref{sec:tokenizer_training}).

The negative impact on downstream task performance was studied on multilingual language models and their tokenizers~\cite{multilingual_tokenizers}. The authors compared dedicated monolingual tokenizers with the multilingual tokenizer from the mBERT model. They concluded that custom tokenizers positively impact the model's downstream performance in nearly all cases.

The second disadvantage to consider is the speed of the models. As pointed out in the recent study~\cite{tokenizer_study}, the out-of-domain tokenizers result in less compressed data (statistically splitting the same text into more tokens) even with significantly larger vocabulary sizes. The inference speed with custom tokenizers then benefits from smaller vocabulary size (smaller input embedding dimension) and more compressed data (fewer tokens produced for the same amount of text). 

Once the tokenizer is trained, it can process the input string, representing the command-line, into a token sequence that can be analyzed by the language model.

\subsection{Model pre-training}
Generally, the performance of language models on downstream tasks benefits from the model learning the language itself first. This phase of model training, called pre-training, utilizes large volumes of data without labels to achieve that goal. There are numerous established pre-training methods in natural language processing, such as masked language modeling introduced in BERT~\cite{bert} or the standard objective for large language models, causal language modeling~\cite{GPT2}. 

For our use case, we have selected the discriminative pre-training objective introduced by ELECTRA~\cite{electra}. The reasons for selecting ELECTRA are twofold:
\begin{itemize}
    \item It is computationally efficient (as demonstrated in the original paper).
    \item The training objective is semantically similar to obfuscation detection.
\end{itemize}
It utilizes two transformer-based language models, a generator, and a discriminator. 
The generator model is trained with the standard masked language modeling objective, where in each sentence, a portion of tokens (e.g., 15\%) is replaced with a [MASK] token, and the model is trained to predict the original identities of the masked-out tokens. The discriminator model is trained to recognize which tokens were replaced by the generator model. As the generator gets better at predicting tokens contextually similar to the original ones, the discriminator has to learn more nuanced differences in the sentence structure. 

After the training, the generator model is discarded, and the discriminator model is fine-tuned for downstream tasks. For more detailed information about the pre-training, refer to the original paper~\cite{electra}.

\subsection{Model Fine-tuning}
In this section, the discriminator model is fine-tuned to detect obfuscated commands. A common issue with detection tasks in the cybersecurity domain is the inherent class imbalance, as malware activities hide in large volumes of benign traffic. Obfuscation detection is no exception. 
From gathered domain knowledge, we estimate the ratio could be as high as one obfuscated sample per hundreds of thousands of execution logs. Such class imbalance makes obfuscation detection an exceptionally hard classification task, as the model must be extremely precise to avoid overwhelming cybersecurity analysts with false positive samples. 

To overcome the class imbalance, we adapt both the fine-tuning dataset, described in Section~\ref{sec:fine-tuning} , and the model training itself. Specifically, we utilized focal loss (FL) introduced in~\cite{focal_loss} for class imbalanced problems. The focal loss for one sample is defined as:

\[ FL(p_t) = −(1 − p_t)^{\gamma}  \log (p_t) \]

\begin{figure}
    \centering
    \includegraphics[width=0.45\textwidth]{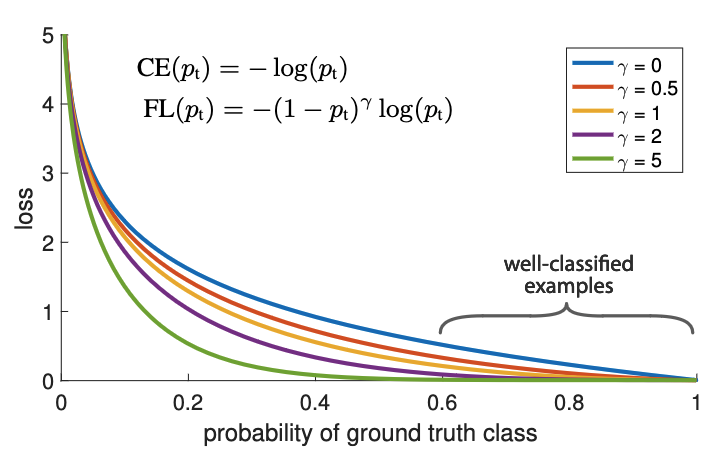}
    \caption{The $\gamma$ parameter controls focus on already well-classified examples. The cross-entropy loss is identical to FL with $\gamma$ set to 0. Image source: \cite{focal_loss}.}
    \label{fig:focal_loss}
\end{figure}

Where $p_t$ is the model's predicted probability of the ground truth label, and $\gamma$ is the focusing hyperparameter of the loss function. The impact of the $\gamma$ parameter and comparison of FL to cross-entropy is displayed in Figure~\ref{fig:focal_loss}. 

The machine learning community already widely adopted FL for class imbalanced problems. For example, in cybersecurity,  FL was successfully utilized in intrusion detection systems \cite{focal_loss_cybersec1, focal_loss_cybersec3}.

\section{Training}
In this section, we provide a detailed description of the training process from Section~\ref{sec:method} using our data. We begin by describing the specific data used, followed by a comprehensive account of each training phase.

\begin{table*}[t]
  \centering
    \caption{Examples of obfuscated samples for each executable. The obfuscation techniques differ based on each binary, the obfuscation for rundll is mainly in obfuscating the name of the DLL library, while cmd and \ps{} allow for more complex obfuscation techniques.}
  \label{tab:obfuscation_examples}
  \begin{tabular}{|c|c|}
   \hline
    \bfseries Binary & \bfseries Obfuscated Command\\
    \hline\hline
    rundll & \begin{minipage}[t]{0.9\textwidth}
            \centering
             \begin{Verbatim}
rundll32.exe  \^&&&^&&^^^&&^^^^^&^&&&&^^&&^^^^&^.{[GUID]},zVh5Hfr3Vd5DLrFl
             \end{Verbatim}
             \end{minipage} \\ \hline
    msiexec & \begin{minipage}[t]{0.9\textwidth}
                \centering
             \begin{Verbatim}
mSIexeC -Q -IhTtP://NT3[.]XyZ:8080/5mGgMqZvXTg/DESKTOP_NAME=USER_NAME
             \end{Verbatim}
             \end{minipage} \\ \hline  
    explorer & \begin{minipage}[t]{0.9\textwidth}
                 \centering
             \begin{Verbatim}
exPLOrER RemoVAbLe Disk
             \end{Verbatim}
             \end{minipage} \\ \hline  
    \ps{} & \begin{minipage}[t]{0.9\textwidth}
             \centering
             \begin{Verbatim}
powershell.exe -exec bypass -noni -nop -w 1 -C IEx( $( set-iTem 'vaRiABLE:OfS' '')+ 
[STrInG]( '91>78!101g11...j116;99'.SpLit('{;gX:<g!j>' ) | forEACH{([int] $_-As[cHAR]) }) + 
$(sET-ItEm 'VaRIaBLe:oFS' ' '))
             \end{Verbatim}
             \end{minipage} \\ \hline  
    cmd & \begin{minipage}[t]{0.9\textwidth}
                 \centering
             \begin{Verbatim}
C:\WINDOWS\system32\cmd.exe /cP^o^w^e^r^S^h^e^l^l^.e^x^e^ -No^Exit
-Ex^ec By^pass -^EC YwBhAG^wAYwA=
             \end{Verbatim}
             \end{minipage} \\ \hline
  \end{tabular}
\end{table*}

\subsection{Data}
While the presented methods are designed to work on top of any source of execution logs, the particular data used in this work were gathered from the Cisco Secure Endpoint (CSE) telemetry. In total, we have gathered approximately 350 million unique execution logs for training and evaluation. This volume corresponds to 20 days of telemetry for $\sim$1000 private networks of various sizes, ranging from universities to industry, finance, or healthcare.

The gathered data contains execution logs for the following LOLBins: \textbf{\ps{}, cmd, msiexec, rundll, and explorer}. An example of an obfuscated command for each selected executable is displayed in Table~\ref{tab:obfuscation_examples}. 

The real-world obfuscated samples for this paper were obtained in cooperation with Cisco Talos. However, because real-world samples are scarce and expensive to acquire, we were able to gather only around 1500 unique real-world samples. Such a low amount of positive samples could present two severe drawbacks. First, a low variety of obfuscation techniques is represented in data for the model to learn and generalize. Second, as mentioned in a study on preprocessing methods for class imbalanced datasets~\cite{imbalanced_data_preprocessing_brabec} and validated in our experiments, with increasing class imbalance, the model training becomes unstable, and models diverge to a majority class classifier, not detecting any obfuscated samples. Therefore, only a limited number of benign samples can be included in the training set.

To avoid these disadvantages, we leveraged two open-source obfuscation tools, Invoke-Obfuscation~\cite{invoke_obfuscation} for \ps{} obfuscation, and Invoke-DOSfuscation~\cite{invoke_dosfuscation} for cmd obfuscation to generate additional "artificially obfuscated" samples. For further details about the artificial samples and utilized obfuscation techniques, please refer to Appendix~\ref{sec:appendix_artificial_samples}.

As each step of model training utilizes a different dataset, the details of their construction are described in each respective section.

\subsection{Tokenizer training}
\label{sec:tokenizer_training}

Training a proper tokenizer is very data intensive. We used 300 million execution logs to train a custom tokenizer with the WordPiece algorithm~\cite{wordpiece} and the Huggingface Tokenizers library~\cite{huggingface}. Since the vocabulary size is a key hyper-parameter for tokenizer training, we have trained tokenizers with various vocabulary sizes: 1k, 5k, 10k, 20k, and 30k to asses the optimal size. 

In the evaluation of tokenizer quality, we focus on two aspects: the data compression and manual inspection of tokenized command-lines. The data compression measures how many tokens each tokenizer produces for the evaluation dataset. In a study on tokenizers~\cite{tokenizer_study}, authors utilize a similar metric, data compression ratio to a baseline tokenizer, to study the impact of various tokenizers on the inference speed. Specifically, the inference speed is influenced by the number of produced tokens and the tokens' embedding dimension equal to the vocabulary size of the utilized tokenizer. 

\begin{figure}[t]
\centering
\includegraphics[width=0.45\textwidth]{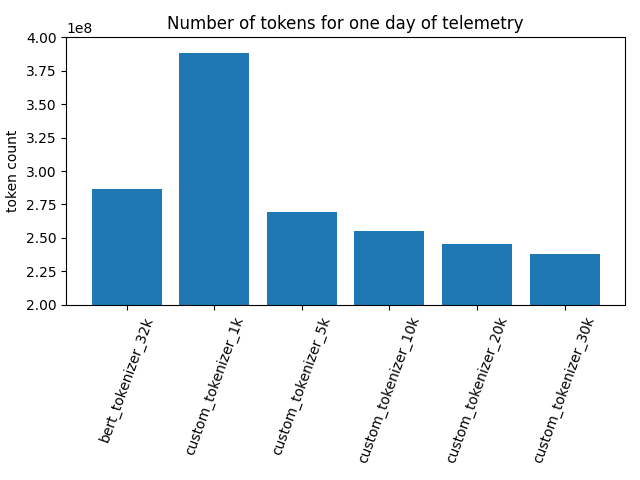}
\caption{\label{fig:token_counts} Comparison of the produced number of tokens for out-of-domain tokenizer and custom tokenizers with various vocabulary sizes on the evaluation dataset.}
\end{figure}

The data compression metric was evaluated on a dataset of never-seen 5 million unique command-lines. In the evaluation, we also included a pre-trained tokenizer with a vocabulary size of 32k trained primarily on natural language (tokenizer from Huggingface hub - \textbf{bert-base-cased}) as a baseline model. The results are displayed in Figure~\ref{fig:token_counts}.

The results show that custom-trained tokenizers achieve significantly better data compression at far smaller vocabulary sizes. For instance, the custom tokenizer with a vocabulary size of 10k produces more than 30 million fewer tokens than the out-of-domain tokenizer while being more than three times smaller. Consequently, the custom tokenizers with sufficient vocabulary sizes offer far better performance from the perspective of inference speed.

\begin{figure*}
    \centering
    \includegraphics[width=0.99\textwidth]{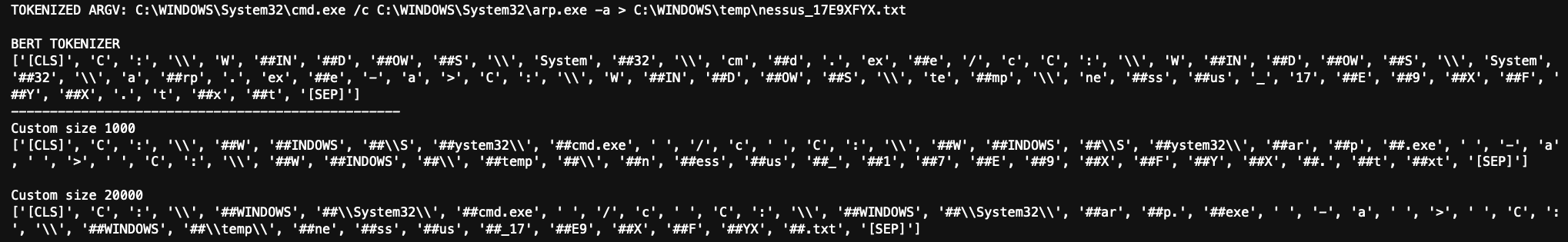}
    \caption{One execution log tokenized by out-of-domain tokenizer (BERT TOKENIZER) and two custom tokenizers for command-line data with vocabulary sizes 1k and 20k, respectively.}
    \label{fig:tokenization_example}
\end{figure*}

To further validate that the custom tokenizers are more suitable for execution logs data, we have manually inspected the produced token sequences on random samples from the evaluation dataset. To illustrate the difference in tokenization, we present an example execution log tokenized by three tokenizers, bert-base-cased, custom\_tokenizer\_1k, and custom\_tokenizer\_20k in Figure~\ref{fig:tokenization_example}. 

The figure clearly illustrates the tokenizers' ability to produce meaningful tokens representing specific patterns in the execution logs, such as ``cmd.exe", ``System32", or ``WINDOWS". As these patterns are prevalent and have an established meaning in the execution logs, we expect the tokenizer to produce one token for each of such patterns.

The analysis led to the following conclusions. The 1k vocabulary size is insufficient for an accurate representation of execution logs, leading to the tokenizer splitting common patterns into arbitrary tokens. Also, the large out-of-domain tokenizer fails to produce meaningful tokens from the prevalent execution log-specific patterns.
The custom tokenizer with a sufficient vocabulary size (20k) produces meaningful tokens representing specific patterns in the execution logs.
From our inspection, the custom tokenizer with a 20k vocabulary size offers the best performance when considering the token quality together with the inference speed and is therefore used for all further experiments.

\subsection{Pre-training}
\label{sec:pretraining_experiments}

\begin{figure}
    \centering
    \includegraphics[width=0.95\linewidth]{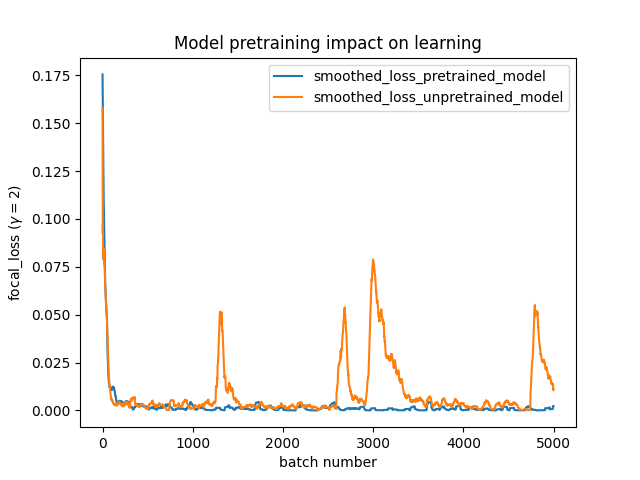}
    \caption{Training loss curves for both pre-trained and non-pre-trained \textit{small} models on the fine-tuning dataset. The non-pre-trained model shows problematic instability during training as compared to the pre-trained model.}
    \label{fig:finetuning_pretrained_unpretrained}
\end{figure}

The ELECTRA pre-training objective is a self-supervised task that does not require labels but benefits from a large amount of data. We used approximately 40 million randomly sampled execution logs to pre-train models with discriminator model sizes (in the number of parameters) 9M, 6M, 4.3M, and 750K, arbitrarily named \textit{large}, \textit{medium}, \textit{small}, and \textit{miniature}. The paired generator models are scaled down by a factor of 4. Since a larger than \textit{small} model did not improve the downstream task performance, the \textit{small} model is used for all further training and experiments.

The details on model parameters and the pre-training process are available in Appendix~\ref{sec:appendix_pretraining}. For our use case, it is vital to analyze whether the pre-training of such a model limited in size has a positive impact on downstream task learning and performance.

The comparison of pre-trained versus non-pre-trained models learning obfuscation detection is displayed in Figure~\ref{fig:finetuning_pretrained_unpretrained}. The figure shows that the pre-trained model is more stable during learning than the non-pre-trained model. The training instability is caused by the nature of the fine-tuning dataset, as discussed in Appendix~\ref{sec:appendix_finetuning}. As the non-pre-trained model also performed worse in evaluation, all further experiments are done with the pre-trained model.

\subsection{Fine-tuning}
\label{sec:fine-tuning}
While the pre-training phase is responsible for getting the model used to the general data, the fine-tuning phase is, in our case, a supervised classification task during which the model learns to classify obfuscated commands. As with most supervised classification tasks, the underlying dataset is a crucial aspect of the problem. The fine-tuning dataset we used has the following structure:
\begin{itemize}
    \item 540k randomly sampled benign data 
    \item 52.5k artificially obfuscated samples
    \item 1.5k real-world obfuscated samples 
\end{itemize}
 The imbalance ratio was set to 10 benign commands per one obfuscated command (real or artificial) as increasing it beyond 30:1 led to training instability and model degradation to the majority classifier. To not further increase the class imbalance and include a wide variety of benign commands simultaneously, the positive class was supplemented by artificially obfuscated samples. The disadvantage of oversampling the positive class with artificial samples is that specific obfuscation techniques present only in real-world samples become sparse in the training data, and the model might not learn to detect them. We validate the model's performance on individual techniques as well as provide details on the training process and caveats in Appendix~\ref{sec:appendix_finetuning}.

The performance of the fine-tuned model on the test dataset, composed of 1/4 of the total samples, is presented in Table \ref{table:finetuned_metrics}.

\begin{table}[!t]
\renewcommand{\arraystretch}{1.3}
\caption{Fine-tuned model's performance on the test dataset.}
\label{table:finetuned_metrics}
\centering
\begin{tabular}{|c|c|c|c|}
\hline
 & \bfseries Precision & \bfseries Recall & \bfseries support\\
\hline\hline
Benign  & 0.9987098 & 0.9999615 & 130045 \\ \hline
Obfuscated & 0.9996846 & 0.9991448 & 12846\\ \hline
\end{tabular}
\end{table}

The test dataset metrics show that the model learned the classification task almost perfectly. As many of the obfuscation techniques significantly impact the structure of the command-line, it was the expected outcome. The different structure is then easily recognizable by an analyst even without an in-depth analysis of the executed commands.

\subsection{Real-world telemetry}

\begin{table}[!t]
\renewcommand{\arraystretch}{1.3}
\caption{Fine-tuned model's evaluation on three days' worth of never-seen data. The detections of the model were manually labeled into three distinct categories. The decision threshold of 0.5 was applied.}
\label{table:production_performance}
\centering
\begin{tabular}{|c|c|}
\hline
 \bfseries Sample category & \bfseries Count \\
\hline\hline
 obfuscated malicious & 38 \\ \hline
obfuscated benign & 148  \\ \hline
non-obfuscated (false positive) & 583  \\ \hline
\end{tabular}
\end{table}

While the model's performance metrics on the test dataset are impressive, it is vital to analyze them with respect to the real-world telemetry, which has class imbalance orders of magnitude higher compared to the test dataset.

Higher class imbalance naturally leads to a significantly lower count of positive samples while the number of produced false positives remains approximately the same. As a result, the precision of the model on real-world telemetry is significantly lower compared to the evaluation on the test dataset.

We can observe the effect when the model is evaluated on three days of never-seen telemetry with a significantly higher class imbalance (our estimate is approximately 1:300 000). Out of 24 million commands, the model assigned a probability of obfuscation greater than 0.1 to approximately 1600 samples. To assess the model's performance, we have manually categorized these samples into the following three categories:

\begin{itemize}
    \item \textbf{obfuscated malicious} - Obfuscated commands produced by either recognized malicious activity or commands with a high probability of being produced by malicious activity worth investigating further.
     \item \textbf{obfuscated benign} - Obfuscated commands produced by non-malicious activity (e.g., excessive use of cmd escape character, using environment variables in the commands, \dots).
    \item \textbf{non-obfuscated}
\end{itemize}

The results for samples with assigned probability higher than the decision threshold of 0.5 are presented in Table~\ref{table:production_performance}. While the model produced approximately the expected amount of false positives, it achieved only around 5\% precision (counting only obfuscated malicious hits as true positives) due to the low number of true positive samples. 

To improve the model's precision on obfuscated malicious commands, we further trained the fine-tuned model to correct its predictions on all the manually labeled samples with an assigned probability of 0.1 or higher. Since analysts are generally also not interested in obfuscated benign samples, the following dataset was constructed for the final training: 
\begin{itemize}
    \item negative class consists of \textbf{non-obfuscated} and \textbf{obfuscated benign} samples
    \item positive class consists of the \textbf{obfuscated malicious} samples supplemented with obfuscated samples from the train portion of the fine-tuning dataset
\end{itemize}
 The supplemented samples from the train portion of the fine-tuning dataset contain all seen obfuscation techniques and were added for two reasons: first, to balance out the high class imbalance to a 1:10 ratio, and second, to prevent the model from overfitting on the 38 positive samples of low variety found during manual labeling. 
 
 In this last phase, the model was trained to correct its predictions on approximately 1700 gathered samples, mostly consisting of manually labeled examples. The resulting model is used in the evaluation section.

\section{Evaluation}
 In this section, we focus on the two most important aspects of the final model: the classification performance and the inference speed.
\begin{figure*}
    \centering
    \includegraphics[width=0.99\linewidth]{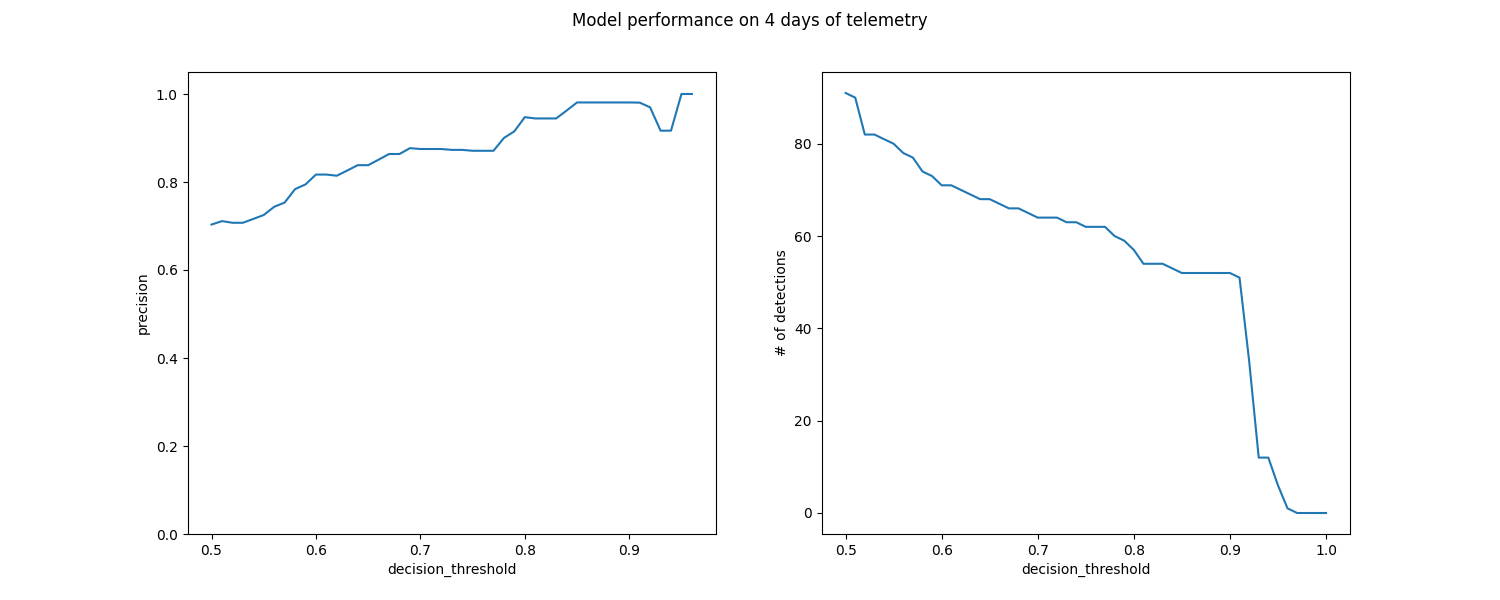}
    \caption{Final model's precision and number of detections based on decision threshold. The metrics were evaluated on four days of never-seen telemetry.}
    \label{fig:inference_precision}
\end{figure*}

\subsection{Classification performance}
\label{sec:model_performance}

To assess the final model's performance, we evaluated the model on 4 more days of new never-seen telemetry and again manually labeled the commands with an assigned obfuscation probability of 0.5 or higher. 

\begin{table}[!t]
\renewcommand{\arraystretch}{1.3}
\caption{Final model's evaluation on four days' worth of never-seen data. The table presents detections of the model manually labeled into three distinct categories.}
\label{table:final_performance}
\centering
\begin{tabular}{|c|c|}
\hline
 \bfseries Sample category & \bfseries Count \\
\hline\hline
 obfuscated malicious & 64 \\ \hline
obfuscated benign & 6  \\ \hline
non-obfuscated (false positive) & 21  \\ \hline
\end{tabular}
\end{table}

The results, presented in Table~\ref{table:final_performance}, show that the last fine-tuning step, where the fine-tuned model adjusted its predictions on a small amount of incorrectly predicted samples, drastically improved the model's precision. Even without raising the decision threshold above 0.5, the model achieved precision above $70\%$.

Furthermore, Figure~\ref{fig:inference_precision} indicates that we can achieve even higher precision by raising the decision threshold. However, the higher precision naturally comes at the cost of losing true positive detections with lower obfuscation probability scores.

\subsection{Inference speed}
The last key parameter of the developed model is the inference speed. We evaluated the model inference speed with the full weights (fp32) and the model weights converted to half-precision (fp16).

\begin{figure}
    \centering
    \includegraphics[width=0.99\linewidth]{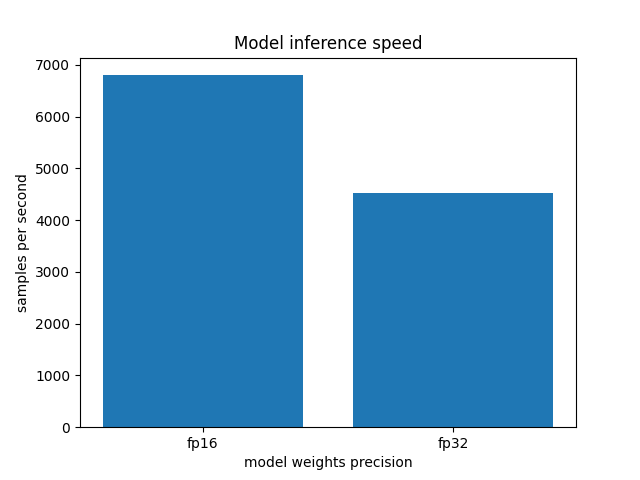}
    \caption{Model inference speed measured on a random sample of one million execution logs. The speed was measured for both the model with full weights (fp32) and with half-precision (fp16) weights. The measurement was done on NVIDIA® T4 16GB GPU. }
    \label{fig:inference_speed}
\end{figure}

Figure \ref{fig:inference_speed} shows that reducing the model weight's precision results in an inference speed gain of $\sim$50\% while no significant classification performance loss was observed compared to the full model. 

Overall, the small model size results in extremely fast inference speed, as the model is capable of analyzing million execution logs in 2.5 minutes even on a single low-end GPU.

\section{Case study}
So far, we presented development steps for a custom NLP-based detection model and completed its evaluation. To further demonstrate the model's value, we present specific examples of obfuscated commands detected by our model that would be missed by common detection approaches for the malware they were attributed to. Specifically, this case study showcases variants of Raspberry Robin and Gamarue, which do not follow previously known patterns. Lastly, we present two new obfuscated samples whose attribution to a malware family or campaign is currently unknown to us.

\subsection{Raspberry Robin}
Raspberry Robin is a worm that leverages multiple pre-installed binaries on the operating system, such as \textit{msiexec}, \textit{rundll32}, \textit{odbcconf}, etc.  Due to the malware's prevalence, it has a well defined kill-chain~\cite{raspberry_robin_onur_blog, raspberry_robin_detection_report}:

\begin{enumerate}
    \item \label{enum:initial_access} initial access through an infected external drive with a malicious \textit{lnk} file
    \item \label{enum:lnk_file_execution} \textit{cmd} executes the malicious \textit{lnk} file
    \item \label{enum:explorer_execution} execution of \textit{explorer} with the infected external drive as the argument
    \item \label{enum:msiexec_install_dll} \textit{msiexec} installs a malicious DLL from a remote location
    \item \label{enum:rundll_bypass} \textit{rundll32} launches the DLL using \textit{odbcconf} or \textit{fodhelper}, etc.
    \item \label{enum:c2_connection} outbound connection to a Command and Control (C\&C) server through TOR
\end{enumerate}

Since the kill-chain is well defined, there are multiple reports~\cite{raspberry_robin_onur_blog, raspberry_robin_detection_report, raspberry_robin_threat_report} that provide signature-like pseudo-code to detect Raspberry Robin. In the context of this case study, we focus only on the signatures that correspond to commands of the kill-chain containing obfuscation techniques.

For instance, none of the reports contain a signature for Step~\ref{enum:lnk_file_execution} of the kill-chain utilizing the white-space insertion technique:

\begin{verbatim}
    C:\Windows\System32\cmd.exe
    \r\r\r\t\r\t\r\r\r\t\r\t\r
    /RCmD<qjM.chK
\end{verbatim}

In this step \textit{cmd} executes a malicious script ``qjM.chK" to which the \textit{lnk} file on the external drive from Step~\ref{enum:initial_access} points to. As the executions from this step usually contain a plethora of combinations, types, and locations of white-space characters, writing a reliable signature for this step is exceptionally difficult. Since the only other reoccurring feature of execution is the commonly used ``/R" switch, the options for a reliable signature-based detector are limited.  However, for our model, the various white-space insertions into the command are obvious signals that the command was obfuscated.

The next phase of the kill-chain, Step~\ref{enum:explorer_execution}, is also reliably detected by our model while no signatures are given by the reports. The execution usually contains an obfuscated execution of the binary \textit{explorer} along with the infected external drive:

\begin{verbatim}
    ExplO^RER USB D^rive
\end{verbatim}

The use of obfuscation techniques, along with changing names of the external drives, results in virtually no overlap among the execution logs of this step in the kill chain. Therefore, the only consistent feature of the executions is the use of obfuscation techniques, which our model detects.

The subsequent of the kill-chain, Step~\ref{enum:msiexec_install_dll}, has a signature-like pseudo-code~\cite{raspberry_robin_detection_report} defined for it:

\begin{itemize}
    \item process == (`msiexec')
    \item process\_command\_line\_includes (`http:', `https')
    \item process\_command\_line\_includes (`-q', `-Q')
\end{itemize}

The signature, however, can be evaded with obfuscation techniques, as can be seen in one of our detections~\cite{raspberry_robin_onur_blog}:

\begin{verbatim}
    mSIexeC -Q-IhtTP://NT3[.]XyZ:8080/
    5qGgVaPvXFg/desktop-name=username
\end{verbatim}

This command is obfuscated using mixed-character case. It may not seem like a significant change, but many signatures expect the protocol to be in its standard lower-case form. Furthermore, the flags ``-Q-I" are concatenated with the URL, which many signatures would not account for. Even such tiny changes are sufficient to avoid detection in many cases.

As demonstrated by the detections observed in our telemetry, we were able to catch numerous steps of the kill-chain, including parts that would be missed by signature detection.

\subsection{Gamarue}
Gamarue, also known as Andromeda or Wauchos, is a worm/botnet consistently ranking among the most prevalent malware families. Due to the malware's prevalence and age, it has a well documented kill-chain~\cite{redcanaryGamarueCanary}:

\begin{enumerate}
    \item \label{enum:gamarue_initial_access} Initial access through an infected external drive with a malicious \textit{lnk} file
    \item \label{enum:gamarue_code_installation} Malicious code installation via \textit{rundll32} referencing the \textit{lnk} file
\end{enumerate}

Similarly, as for Raspberry Robin, there is a signature-like pseudo-code defined for the execution of Step~\ref{enum:gamarue_code_installation} of the kill-chain:

\begin{itemize}
    \item process == (`rundll32')
    \item process\_command\_line\_includes (`/\textbackslash S\{10, 70\}.\textbackslash S\{10, 70\},\textbackslash w\{16\}/')
\end{itemize}

The signature already accounts for obfuscation since the arguments of \textit{rundll32} use uncommon character combinations changing in each execution. However, even slight modifications of the command by changing the counts of letters from the expected 10-70 characters to a lower count enough to evade detection, as can be seen in the sample detected by our model:

\begin{verbatim}
    C:\Windows\System32\rundll32.exe  \BrP
    5lzjT.P5d.RvTxRB.zfLtZJ.BjT.1lV.vbHl.l
    VFzl,cUySwoImK8cAeW4Y
\end{verbatim}

This again demonstrates that our obfuscation detection model can detect novel samples even for established malware with well documented kill-chain.

\subsection{Unknown malware obfuscation-based detections}
Our model's biggest strength, however, lies in detecting unknown/unseen obfuscated commands. We present two novel examples it has detected to demonstrate the model's capabilities.
 
 The first example contains two hard-to-detect techniques: variable names obfuscated by using ``-", ``\#", etc. in their names and constructing the command from these variables during execution as can be seen in the following command:
 
\begin{verbatim}
    cmd.exe /c set --$#$--=net&&
    set '''=at&&set ;;;;=st&&
    cmd.exe /c %--$#$--%%;;;;%%'''
    % -s -p UDP 
\end{verbatim}

The goal of this command is to check statistics of UDP traffic using \textit{netstat}, which could be used as a part of the initial discovery process in a compromised environment. 

Another example of an obfuscated command flagged by our model is the following \textit{\ps{}} execution:

\begin{verbatim}
    C:\Windows\System32\WindowsPowerShell\
    v1.0\powershell.exe -w hidden $test= 
    'htt'+'p://'+'XX[.]XX[.]XX[.]XX'+
    '/login.php';$response = 
    $(New-Objectsystem.Net.WebClient)
    .DownloadString($test);
    $log=$response;
    $command =[scriptblock]::Create($log);
    &$command; 
\end{verbatim}

The \textit{\ps{}} sample is obfuscated by splitting the original command into multiple strings and concatenating their content during execution. Specifically, this command downloads and executes the content from a URL with an embedded IP address: ``http://XX[.]XX[.]XX[.]XX/login.php". The contents of the executed script are currently unknown since the website was not available at the time of writing this paper. As this is a common technique used by malware such as Metamorfo (or Casbaniero)~\cite{metamorfo}, the command is highly suspicious and likely malicious.

\section{Conclusion}

In this paper, we presented a novel method for detecting obfuscated command-lines using custom-trained small transformer neural networks. We provided a detailed step-by-step description of training such a model from scratch. Notably, the NLP-based approach allowed us to extend the model's detection capabilities beyond \ps{} to multiple other LOLBins, which is a significant advancement compared to previous works on obfuscation detection.
By evaluating numerous days of real-world telemetry, we have demonstrated that the model produces high-precision detections even for large telemetries. The practical value of our model was further highlighted in a case study of two prevalent malware families that heavily utilize obfuscation in various parts of their lifecycles. The model detected novel samples that did not follow previously known patterns. As future work, we are extending our detection capabilities from Windows-based execution logs to Unix-based systems.

\subsection*{Acknowledgments}

We would like to thank all cybersecurity analysts (Cisco Talos team and Treo team) who provided us with much-needed real-world obfuscation samples for our research and were instrumental in analyzing obfuscated samples with a special thanks to Meghan Correa from Cisco Talos for her exceptional contributions and cooperation. We would also like to express our gratitude to Jaroslav Hlavac, Benjamin Paterek, and Petr Pulc for their insightful discussions and comments.

\bibliographystyle{ieeetr}
\bibliography{obfuscation}

\appendices

\section{Artificially obfuscated samples}
\label{sec:appendix_artificial_samples}

\begin{figure*}[!ht]
    \centering
    \includegraphics[width=0.7\linewidth]{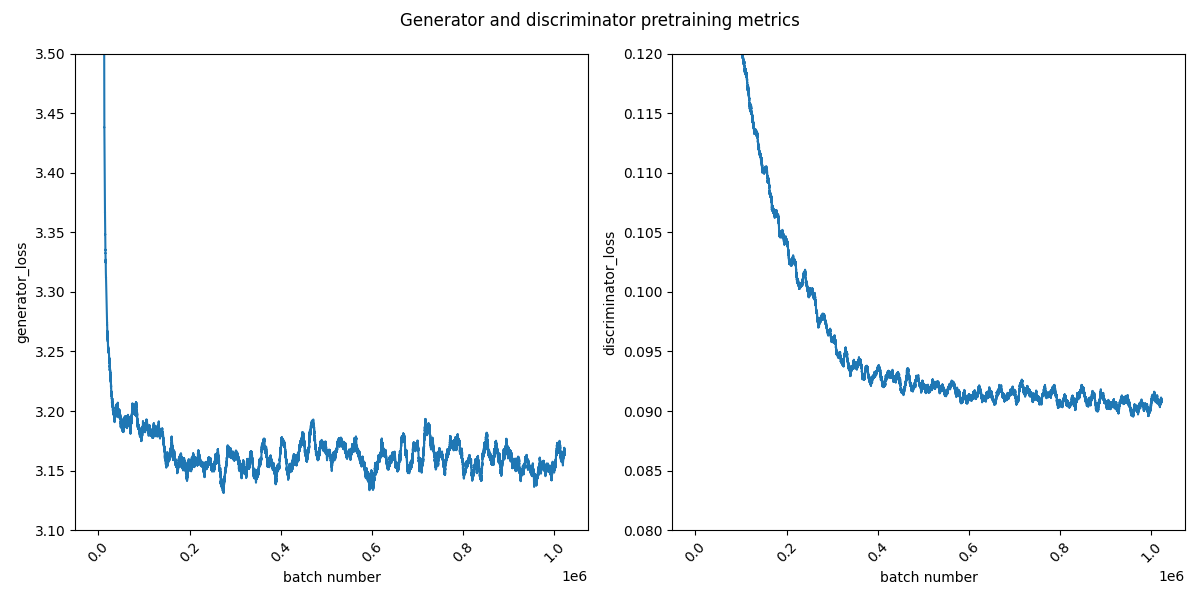}
    \caption{Pre-training loss curves for the \textit{small} model size. The generator stops learning early in the training, limited by its smaller size. The discriminator stops learning shortly after.}
    \label{fig:pretrianing_small_model}
\end{figure*}

\begin{figure*}[t]
\centering
\includegraphics[width=0.7\textwidth]{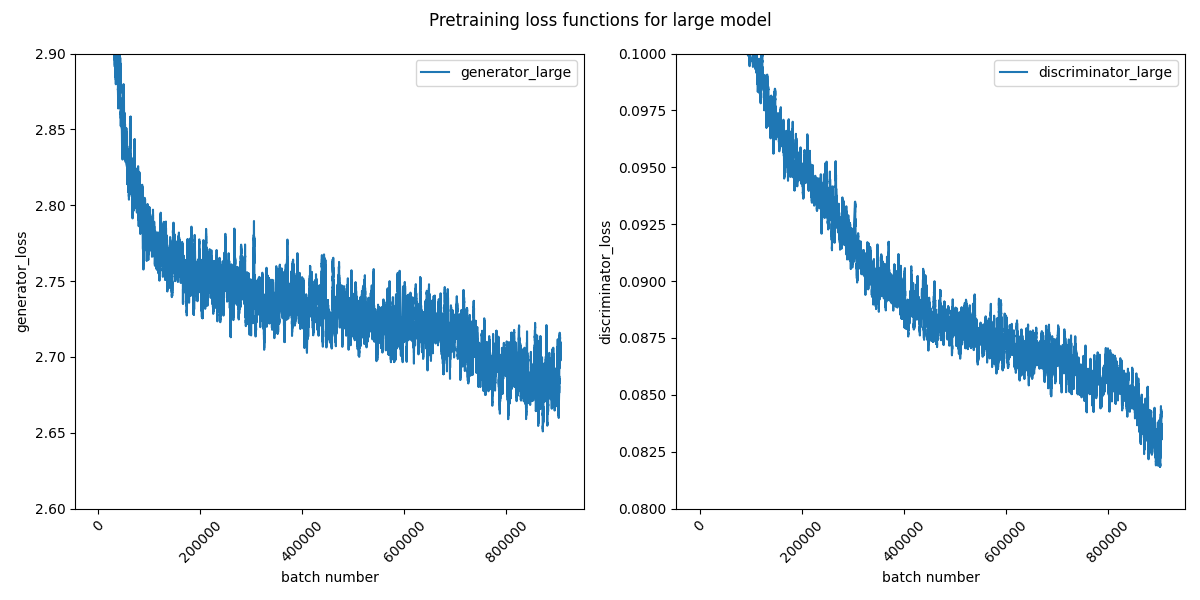}
\caption{\label{fig:pretraining_large} Large model pre-training metrics. Both the generator and discriminator models continue learning throughout the entire learning process.}
\end{figure*}

This section describes the process of creating the artificially obfuscated samples for the fine-tuning dataset from the gathered execution logs. Generally, the execution log consists of two parts. The first part specifies the executed LOLBin, while the second represents the executable's arguments. For example, in the following execution log:
\begin{verbatim}
    C:\WINDOWS\system32\cmd.exe /c 
    tasklist.exe /fi imagename eq logonui* 
    /fi session eq 11,684
\end{verbatim}
the executed binary is \begin{verbatim}
    C:\WINDOWS\system32\cmd.exe
\end{verbatim}
 and the actual executed cmd command given by the arguments follows: \begin{verbatim}
    /c tasklist.exe /fi imagename eq 
    logonui* /fi session eq 11,684
\end{verbatim}
As the executed binary is supplied by the operating system, we apply obfuscation techniques to the actual command only.
The command is passed down to the respective obfuscation tool, either Invoke-Obfuscation for \ps{} or Invoke-DOSfuscation for cmd. 

We have selected a set of obfuscation techniques to be applied to a random sample of extracted commands. However, some techniques apply only to commands containing specific elements, such as strings, and there is no guarantee that a randomly sampled command would contain these elements. In such cases, the commands would not be obfuscated and would introduce wrongly labeled samples in our dataset. To minimize such occurrences, we have tried to select/combine techniques that apply to a more general set of commands. Utilized obfuscation techniques are listed in Table \ref{table:obfuscation_techniques}.

\begin{table}[!t]
\renewcommand{\arraystretch}{1.3}
\caption{Utilized obfuscation techniques.}
\label{table:obfuscation_techniques}
\centering
\begin{tabular}{|c|c|c|}
\hline
 \bfseries executable & \bfseries obfuscation technique & \bfseries tool command  \\
\hline\hline
\multirow{12}{*}{\centering powershell} & token\_obfuscation & token,all,1 \\ \cline{2-3}
& command\_compressing & compress,1 \\ \cline{2-3}
& command\_encoding\_ASCII & encoding,1 \\ \cline{2-3}
& command\_encoding\_hex & encoding,2  \\ \cline{2-3}
& command\_encoding\_octal & encoding,3  \\ \cline{2-3}
& command\_encoding\_binary & encoding,4  \\ \cline{2-3}
& command\_encoding\_SecureString & encoding,5  \\ \cline{2-3}
& command\_encoding\_BXOR & encoding,6  \\ \cline{2-3}
& command\_encoding\_special\_chars & encoding,7  \\ \cline{2-3}
& command\_encoding\_whitespace & encoding,8  \\ \cline{2-3}
& command\_string\_concatenate & string,1 \\ \cline{2-3}
& command\_string\_concatenate\_reorder & string,2 \\ \hline
\hline
\multirow{7}{*}{cmd} & environment\_variable\_light & encoding,1 \\ \cline{2-3}
& environment\_variable\_medium & encoding,2 \\ \cline{2-3}
 & payload\_concat\_light & payload,concat,1 \\ \cline{2-3}
 & payload\_concat\_medium & payload,concat,2 \\ \cline{2-3}
 & payload\_concat\_reverse\_light & payload,reverse,1 \\ \cline{2-3}
 & payload\_concat\_reverse\_medium & payload,reverse,2 \\ \cline{2-3}
 & payload\_forcode & payload,forcode,1 \\ \hline
\end{tabular}
\end{table}

Each listed technique is applied to a sample of random commands from the telemetry with the respective obfuscation tool so that all listed techniques are equally represented in the fine-tuning dataset.  When calling each obfuscation tool to obfuscate commands from \ps{} console, the ``tool command" column contains a string that specifies which technique the tool should apply\footnote{When specifying the technique in the interactive console of the obfuscation tool, the commas have to be replaced by forward slashes.}.  In the case of Invoke-Obfuscation, the command for applying the ``token\_obfuscation" technique to an extracted command ``command-to-obfuscate"  would result in:
\begin{verbatim}
  Invoke-Obfuscation -ScriptBlock 
  {command-to-obfuscate} -Command 
  'token,all,1' -Quiet
\end{verbatim}

Similarly, for Invoke-DOSfuscation and the technique ``environment\_variable\_light":
\begin{verbatim}
  Invoke-DOSfuscation -Command 
  command-to-obfuscate -CliCommand 
  "encoding,1" -Quiet;
\end{verbatim}

\section{Pre-training details}
\label{sec:appendix_pretraining}

\begin{table*}[t]
  \centering
  \caption{Parameter specifications for each model size.}
  \label{table:pretraining_parameter_specifications}
  \begin{tabular}{|c|c|c|c|c|c|c|}
   \hline
    \bfseries size & \bfseries model &  \bfseries embedding\_size & \bfseries hidden\_size & \bfseries num\_hidden\_layers & \bfseries intermediate\_size & \bfseries num\_attention\_heads \\
    \hline \hline

    \multirow{2}{*}{\centering large} & discriminator & 128 & 256 & 8 & 1024 & 4 \\ \cline{2-7}
     & generator & 128 & 64 & 8 & 256 & 1 \\ \hline

     \multirow{2}{*}{\centering medium} & discriminator & 128 & 256 & 4 & 1024 & 4 \\ \cline{2-7}
     & generator & 128 & 64 & 4 & 256 & 1 \\ \hline

     \multirow{2}{*}{\centering small} & discriminator & 128 & 256 & 2 & 1024 & 4 \\ \cline{2-7}
     & generator & 128 & 64 & 2 & 256 & 1 \\ \hline

     \multirow{2}{*}{\centering miniature} & discriminator & 32 & 64 & 2 & 256 & 2 \\ \cline{2-7}
     & generator & 16 & 32 & 2 & 64 & 1 \\ \hline
  \end{tabular}
\end{table*}

The parameter configurations for each model size are listed in Table~\ref{table:pretraining_parameter_specifications}. The configurations were inspired by the paper introducing the ELECTRA style of pre-training. Training hyper-parameters were copied from the study for all training runs and are listed in Table~\ref{table:pretraining_parameters}

\begin{table}[!h]
\caption{Pre-training hyper-parameters}
\label{table:pretraining_parameters}
\centering
\begin{tabular}{|c|c|}
\hline
 \bfseries Hyper-parameter & \bfseries value \\
\hline\hline
 learning rate & 5e-4 \\ \hline
 weight decay & 0.01  \\ \hline
 batch size & 32  \\ \hline
\end{tabular}
\end{table}

Model pre-training was performed for every model size and custom tokenizer with every vocabulary size. The impact of vocabulary size on pre-training was consistent with the tokenization evaluation in Section~\ref{sec:tokenizer_training}, where the best performance provided custom tokenizers with sizes 10k and higher. To be consistent with the rest of the paper, all further described experiments utilize the custom tokenizer with a 20k vocabulary size. 

However, tokenizer quality is only one of the key factors in model pre-training. A recent large-scale study on pre-training language models shows that pre-training dominantly depends on the model size \cite{llm_scaling}. We observed this effect in our model pre-training as well. 

Figure~\ref{fig:pretrianing_small_model} shows that the \textit{small} model could achieve only limited performance during pre-training. Both the generator and discriminator learning are halted early in the training process. However, the \textit{medium} and \textit{large} model sizes were far less limited in this regard. The pre-training metrics for the \textit{large} model are depicted in Figure~\ref{fig:pretraining_large}. The figure shows that the larger model size allowed the generator to continually learn throughout the entire learning process. Furthermore, the discriminator learning was also not halted and continued throughout the entire pre-training duration, achieving significantly lower loss function values. 

While the \textit{medium} and \textit{large} models offer better performance with regards to pre-training metrics and could be used for advanced tasks such as semantic similarity search, we observed little to no advantage over the pre-trained \textit{small} model in the fine-tuning phase. 

\section{Model fine-tuning}
\label{sec:appendix_finetuning}

\begin{figure}[h]
    \centering
    \includegraphics[width=0.99\linewidth]{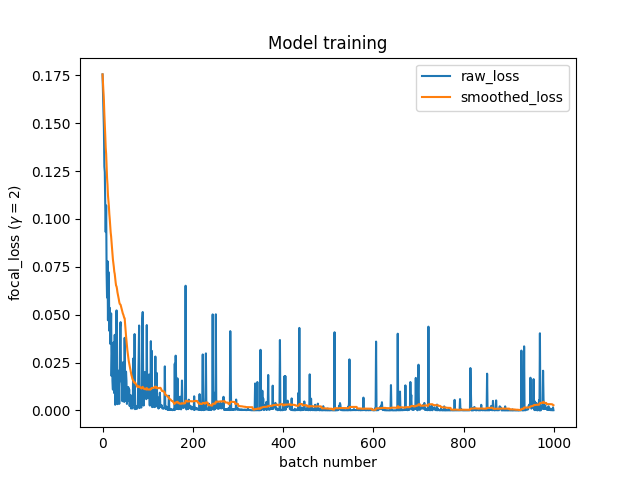}
    \caption{Fine-tuning loss curve of the pre-trained model.}
    \label{fig:model_finetuning}
\end{figure}

\begin{figure*}[t]
    \centering
    \includegraphics[width=0.99\textwidth]{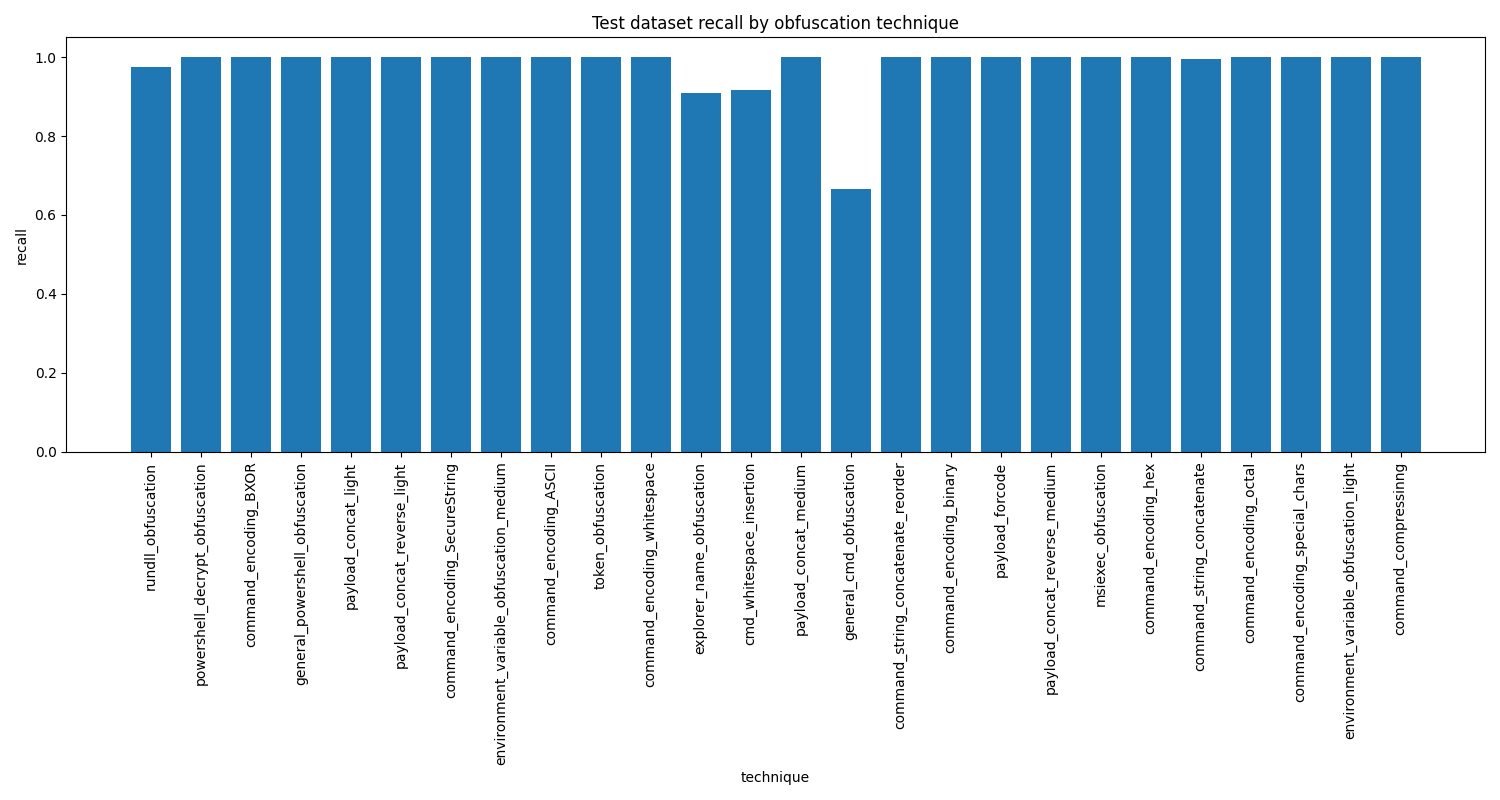}
    \caption{Test dataset recall for individual obfuscation techniques.}
    \label{fig:test_recall_techniques}
\end{figure*}

The fine-tuning dataset contains samples that inherently cause the model training to be unstable. The fine-tuning loss curve for the pre-trained model is displayed in Figure~\ref{fig:model_finetuning}. The model starts learning the task quickly, as expected. However, the loss function peaks occurring during training indicate inconsistencies in the dataset. Closer inspection of the samples causing loss function peaks reveals that three types of samples are responsible for confusing the model:

\begin{itemize}
    \item incorrectly obfuscated (by the obfuscation tool) commands labeled as obfuscated 
    \item obfuscated commands in the sampled telemetry automatically labeled as benign
    \item obfuscated samples with rare obfuscation technique
\end{itemize}
Since it is not feasible to manually label and inspect every artificially obfuscated command at the scale of the dataset, there are rare samples where the obfuscation tool fails at obfuscating the command. A common case is when a certain obfuscation technique does not apply to the particular sampled command (e.g., no string token in a \ps{} command to obfuscate), as discussed in Appendix~\ref{sec:appendix_artificial_samples}. Such samples were easily identified by the loss peak analysis and removed from the dataset.

As for the obfuscated samples in telemetry, they very rarely come from malware activity. However, some benign commands excessively use techniques heavily utilized in obfuscation, such as escaping characters, command encoding, etc. These benign commands cause a large portion of the training peaks, as they are not that uncommon. 

\newpage

An example of such a command excessively using command-line escaping characters is:
\begin{verbatim}
C:\Windows\system32\cmd.exe /d /s 
/c mocha ^^^--recursive^^^ 
^^^--colors^^^ ^^^./test/config.js^^^
^^^./test^^^ ^^^--exit^^^
\end{verbatim}
Removing the commands from the telemetry is not an option, since the model should not trigger on these samples to keep high precision of interesting detections. These samples are structurally very close to some obfuscation techniques, such as character insertion, and as a result, the model experiences large loss function peaks by confidently predicting these samples as obfuscated. While the pattern is similar to some obfuscation techniques, the model learned to distinguish between these samples and actual obfuscation techniques during the last phase of model training described in Section~\ref{sec:fine-tuning}.

The last category consists of samples from real telemetry utilizing techniques unavailable in the obfuscation tools. Since such samples are exceedingly rare in the fine-tuning dataset, the classification model tends to focus on the more prevalent techniques.

The model's recall for each obfuscation technique in the test dataset is depicted in Figure~\ref{fig:test_recall_techniques}. The figure shows that the model's performance on some obfuscation techniques present only in real-world samples, such as ``explorer\_name\_obfuscation" or ``cmd\_whitespace\_insertion" is slightly lower as there are only a few samples for these techniques in the dataset. 

\end{document}